\pdfoutput=1

\documentclass[11pt]{article}

\usepackage{naacl2021}

\usepackage{times}
\usepackage{latexsym}

\usepackage{fancyvrb}
\usepackage{amsmath}
\usepackage{graphicx}
\usepackage{amssymb}
\usepackage{url}
\usepackage{cprotect}

\usepackage[T1]{fontenc}

\usepackage[utf8]{inputenc}

\usepackage{microtype}

\title{A Simple Approach for Handling Out-of-Vocabulary Identifiers \\ in Deep Learning for Source Code}

\author{Nadezhda Chirkova\thanks{\ \ The work was done while working at Samsung-HSE Laboratory, HSE University. Both authors contributed equally. } \and Sergey Troshin$^{*}$ \\
  HSE University \\
  Moscow, Russia \\
  \texttt{\{nchirkova, stroshin\}@hse.ru}}

\begin{document}
\maketitle
\begin{abstract}
There is an emerging interest in the application of natural language processing models to source code processing tasks. One of the major problems in applying deep learning  to software engineering is that source code often contains a lot of rare identifiers, resulting in huge vocabularies.
We propose a simple, yet effective method, based on identifier anonymization, to handle out-of-vocabulary (OOV) identifiers. Our method can be treated as a preprocessing step and, therefore, allows for easy implementation. We show that the proposed OOV anonymization method significantly improves the performance of the Transformer in two code processing tasks: code completion and bug fixing. 
\end{abstract}

\section{Introduction}
Natural language processing (NLP) is widely used for source code processing (SCP), e.\,g.\,for learning the meaningful vector representations of code \cite{codebert, code2vec, user2code2vec}, that can be used in various downstream tasks, e.\,g.\,code summarization \cite{sode_summarization_neural_attention, tree_encoding}, code completion 
\cite{code-prediction-transformer}, or bug fixing~\cite{Hellendoorn}.

An important question, one should answer before building an SCP model, is how to create a vocabulary? \citet{code_bpe} underline that modern source code datasets may incorporate millions of unique identifiers, of which less than 1\% occur in the dataset frequently, e.\,g.\,more than 5 times. The common practice is to crop the vocabulary based on top-N identifiers and replace all occurrences of out-of-vocabulary (OOV) identifiers with an \verb|UNK| identifier to avoid huge embedding matrices and the meaningless embeddings of rare tokens. But can one process rare identifiers in a better way?

\begin{figure}[t!]
    \centering
        \centering
                \begin{tabular}{l}
                Vocabulary: $\{$\verb|np|, \,\,\verb|sin| $\}$ \\
                Input: \verb|my_y  = np.sin(my_x) + my_x| \\
                Standard OOV processing procedure:  \\
                \hspace{1.83cm} \verb|UNK = np.sin(UNK) + UNK| \\
                Proposed OOV anonymization procedure: \\
                \hspace{1.17cm} \verb|VAR1 = np.sin(VAR2) + VAR2|
                \end{tabular}
        \cprotect\caption{Illustration of the proposed OOV anonymization procedure. Out-of-vocabulary identifiers \verb|my_y| and \verb|my_x| are replaced with anonymized identifiers \verb|VAR1| and \verb|VAR2|, while in-vocabulary identifiers \verb|np| and \verb|sin| preserve their names.}
        \label{fig:ill}
\end{figure}

\begin{figure}[t!]
    \centering
        \centering
        \begin{tabular}{c}
               Python150k dataset (custom train-test split)\\
                \includegraphics[width=0.97\linewidth]{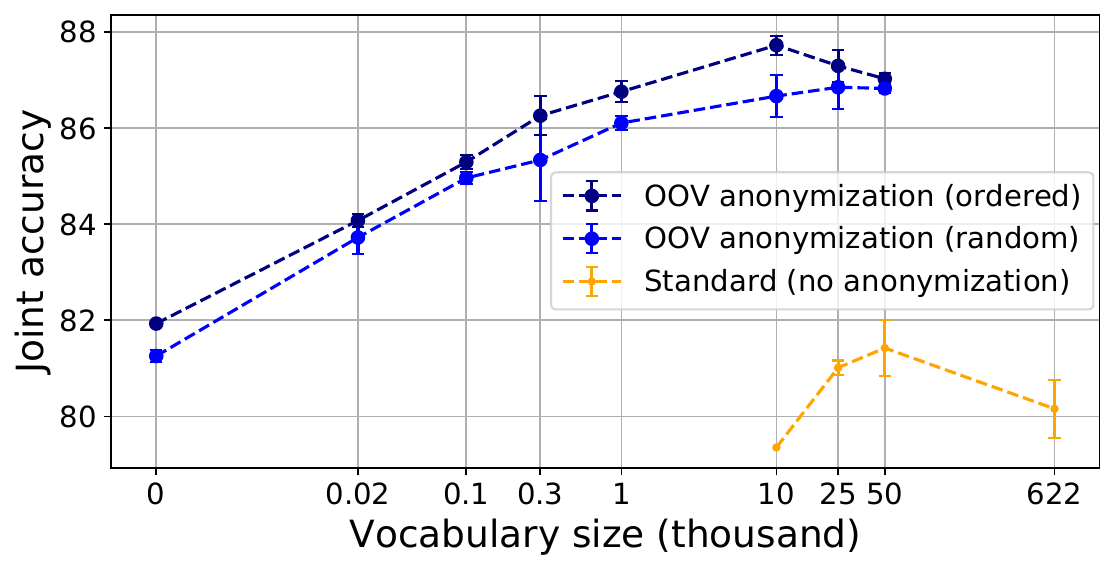} \\
                JavaScript150k dataset (custom train-test split) \\
                \includegraphics[width=0.97\linewidth]{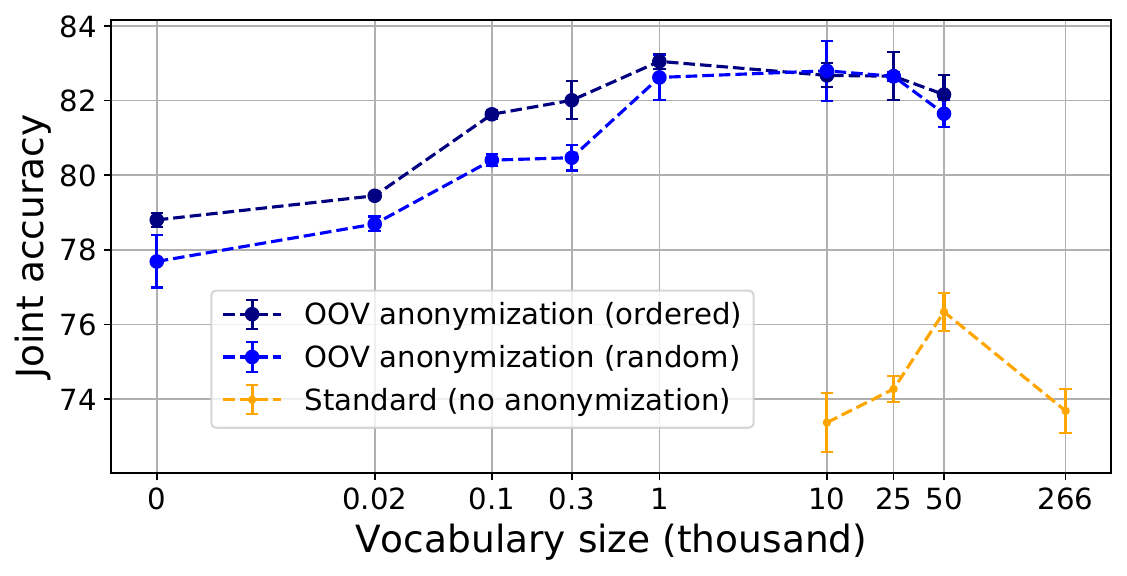}
                \end{tabular}
        \cprotect\caption{Results for Transformer in the variable misuse task: joint bug localization and repair accuracy, mean value $\pm$ standard deviation (over 3 runs). Models with the proposed OOV anonymization 
        significantly outperform the standard model (all OOV identifiers are replaced with an \verb|UNK| token). The numerical data for the plots is given in Table~\ref{tab:results_all_vm} in Appendix.
        }
        \label{fig:res_varmisuse}
\end{figure}

There are two main directions in the NLP literature to tackle rare tokens: open vocabulary and copy-based approaches.
An open vocabulary solution implies splitting rare tokens into subtokens \cite{bpe_nlp}. 
The copy-based approaches are used in generation tasks and imply using
the pointer mechanism \cite{pointer_nlp} to copy tokens from the input sequence.

We propose a new, simple, yet effective approach for processing OOV identifiers in source code, namely OOV anonymization. Anonymization implies replacing \textit{rare} identifiers with unique placeholders, i.\,e.\,\verb|VAR1|, \verb|VAR2|, \verb|VAR3| etc., while preserving the names of \textit{frequent} identifiers. An example of OOV anonymization is shown in figure~\ref{fig:ill}. The intuition behind using anonymization is that it preserves the semantics of the algorithm that the code snippet implements, i.\,e.\, renaming user-defined identifiers does not change the underlying algorithm. By contrast, replacing all rare identifiers with an \verb|UNK| identifier changes the algorithm. We underline that we propose anonymizing only rare identifiers, because frequently used identifier names may serve as an additional source of information, and neural networks are indeed capable of capturing this information.

The proposed OOV anonymization strategy allows for easy implementation as a preprocessing step, thus no modification of model code is required. Another advantage of the OOV anonymization is that it enhances both the encoder  and the decoder. The proposed approach significantly outperforms the model with all rare identifiers being replaced with an \verb|UNK|, in code completion and bug fixing tasks, with the Transformer~\cite{vaswani_transformer} architecture being used (see example comparison in Fig.~\ref{fig:res_varmisuse}). Our code and data split are available at \url{https://github.com/bayesgroup/code_transformers}.

\section{Related Work}
\label{related}

\paragraph{Handling OOV identifiers in source code.} Code processing often borrows ideas from NLP. Source code can be represented as a sequence of identifiers. In this case, identifiers can be further split into subtokens using byte-pair encoding (BPE) \citep{code_bpe, bpe_nlp} resulting in an open vocabulary model. This approach has several drawbacks. Firstly, splitting identifiers into subtokens increases the length of the sequence several times. This substantially slows down inference, e.\,g.\, vanilla Transformer's forward pass has a complexity quadratic w.\,r.\,t.\,the input length. Secondly, splitting breaks one-to-one alignment between identifiers and nodes in the parsing tree, e.\,g.\,abstract syntax tree (AST), in other words, \textit{several} subtokens correspond to \textit{one} node in the AST, which makes it harder to apply structure-aware models such as~\cite{Hellendoorn} or \cite{code2seq}. 
To the best of our knowledge, all SCP works, proposing structure-aware models, either use entire tokens without subtokenization~/~BPE, or average the embeddings over subtokens (this strategy provides only a slight quality improvement compared to the first one), and the question of how to incorporate BPE in structure-aware models needs further investigation. Taking into account the described disadvantages of BPE, we do not consider BPE in this work and do not split tokens into subtokens.

An orthogonal direction for handling OOV identifiers in source code is the modification of the computational graph. For the task of code generation,
the pointer mechanism is widely adapted~\cite{pointer}. \citet{open_vocabulary_graph_cache} also propose a graph-structured cache for inferring the representations of the rare identifiers in source code. The major drawback of the mentioned approaches is that they are quite hard to implement. 

\paragraph{Identifier anonymization in source code.}
\citet{chirkova2020empirical} conduct an empirical study of Transformers for source code in a setting with \textit{all} identifiers being anonymized and show that Transformers can make meaningful predictions in this setting. By contrast, we propose anonymizing only OOV identifiers and show that it boosts the performance of the model in the setting with frequent identifier names being present in the data. The anonymization of \textit{all} identifiers has also been used in \cite{deepfix} and \cite{commitgen} for training recurrent neural networks. \citet{student_code} replace variables with their types, losing information about identifier repetition.

\section{Proposed method}
\label{ours}
    Consider a vocabulary of all identifiers in the training data. It could be a vocabulary of all tokens if we treat input code snippets as text sequences, or a vocabulary of all user-defined variables if we parse the ASTs of code snippets. Let us now select the vocabulary $V_{full}$ of frequent identifiers and call all others OOV identifiers.
    
    We propose an elegant way of tackling OOV identifiers
    based on anonymization. Particularly, we propose replacing all OOV identifiers with placeholders \verb|VAR1|, \verb|VAR2|, \verb|VAR3| etc.  All occurrences of one identifier in one input sequence are replaced with the same placeholder (anonymized identifier), but different identifiers are replaced with different placeholders. One identifier may be replaced with different placeholders in different input sequences. An example of OOV anonymization is presented in figure~\ref{fig:ill}.
    
    We consider two strategies for the OOV anonymization, namely \textit{ordered} anonymization and \textit{randomized} anonymization. The \textit{ordered} anonymization implies assigning an anonymized identifier \verb|VAR1| to the first seen rare identifier, \verb|VAR2| to the next seen rare identifier, etc.  For example, the snippet from Fig.~\ref{fig:ill} is transformed into \verb|VAR1 = np.sin(VAR2) + VAR2|.
    The \textit{randomized} anonymization implies 
     fixing the placeholder vocabulary size $|V_{an}|$ and selecting a random subset of anonymized placeholders 
     $\verb|VAR1|$, $\dots$, \verb|VAR|$|$\verb|V|${_{an}}|$
     for each code snippet. For example, the snippet from Fig.~\ref{fig:ill} can be transformed into \verb|VAR38 = np.sin(VAR801) + VAR801|. To ensure that we can always 
 encode identifiers in a code snippet injectively, the size $|V_{an}|$ of the placeholder vocabulary should not be fewer than the maximum possible number of tokens per snippet. We set $|V_{an}|$ to the maximum length of code snippets.
    
    The proposed OOV anonymization can be seen as a preprocessing step, thus no model parts change. In the encoder, the embedding matrix contains embeddings for both anonymized and in-vocabulary identifiers: $\{e_v\}_{v \in V_{full}} \cup \{e_{\verb|VARi|}\}_{i=1}^{|V_{an}|}$. In the decoder, when generating the next identifier, the softmax is computed over all anonymized and in-vocabulary identifiers. We note that the \textit{ordered} OOV anonymization may need a more careful implementation, e.\,g.\,of metric computation, see details in section~\ref{sec:experiments}.

\section{Experiments}
\label{sec:experiments}

\subsection{Experimental setup}
We conduct experiments with Transformer~\cite{vaswani_transformer} on the code completion (CC) and variable misuse (VM) tasks, on Python150k~\cite{python150k} (the redistributable version of~\cite{redistr}) and JavaScript150k~\cite{javascript150k} datasets. 

We use the problem setup, metrics and loss of~\citet{Hellendoorn} for the VM task, and of~\citet{code-prediction-transformer} for the CC task. To validate our implementation, we check that the quality we obtain with the vanilla Transformer is the same as the quality of this model reported in the corresponding works, see details in Appendix~\ref{app:reproduce}. As a base model, we use the 6-layer Transformer equipped with the relative attention mechanism~\cite{seq_rel_attn} and applied over the depth-first traversal of the AST. \citet{chirkova2020empirical} show that such an approach leads to high performance and outperforms the vanilla Transformer and several techniques for capturing AST structure in Transformer. The hyperparameters 
are given in Appendix~\ref{sec:appendix_details}.
\citet{allamanis, chirkova2020empirical}  emphasize the importance of the thoughtful splitting data into training and testing parts, which includes \textit{splitting by repositories} and \textit{removing duplicate code}. We follow the same strategy in our experiments (later referred to as custom train-test split). 

\textbf{Variable misuse task.\quad} 
For the VM task, we use the same setup as in~\citep{Hellendoorn}, below we briefly recap this setup.
In the VM task, given the code of a function, the task is to output two positions (using two pointers): in what position a wrong variable is used and which position a correct variable can be copied from (any such position is accepted). If a snippet is non-buggy, the first pointer should select a special no-bug position. We obtain two pointers by applying two position-wise fully-connected layers, and softmax over positions on top of the Transformer encoder outputs. We use the joint 
accuracy to assess the model quality (the portion of buggy examples for which the model correctly localizes and repairs the bug).

To obtain a dataset for the VM task, we select all top-level functions in Python150k dataset, including functions inside classes, and filter out functions longer than 250 AST nodes, and functions with less than three  positions containing user-defined variables or less than three distinct user-defined variables. The resulting training~/~testing set consists of 417K~/~231K functions (Python) and 202K~/~108K functions (JavaScript). One function may occur in the dataset up to 6 times, 3 times with synthetically generated bug and 3 times without bug. The buggy examples are generated synthetically by choosing random bug and fix positions from positions containing user-defined variables. When using the ordered OOV anonymization, we firstly inject a synthetic bug and then perform anonymization, to avoid data leak.

\begin{figure}[t]
    \centering
    ~~~~~~Python150k dataset (custom train-test split) \\
         \includegraphics[width=\linewidth]{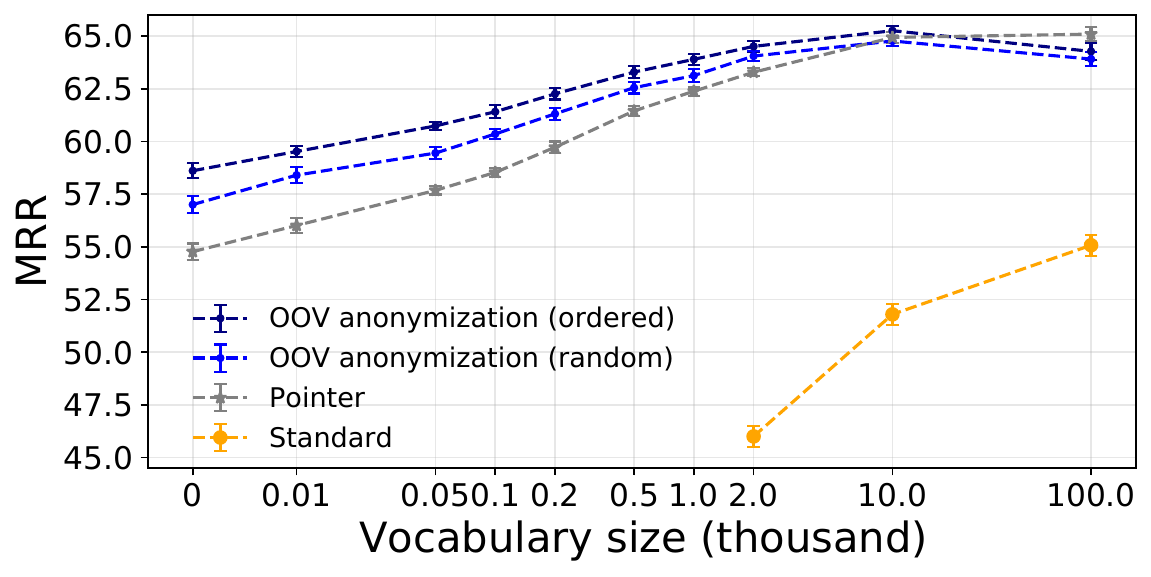}  \\
         ~~~JavaScript150k dataset (custom train-test split) \\
         \includegraphics[width=\linewidth]{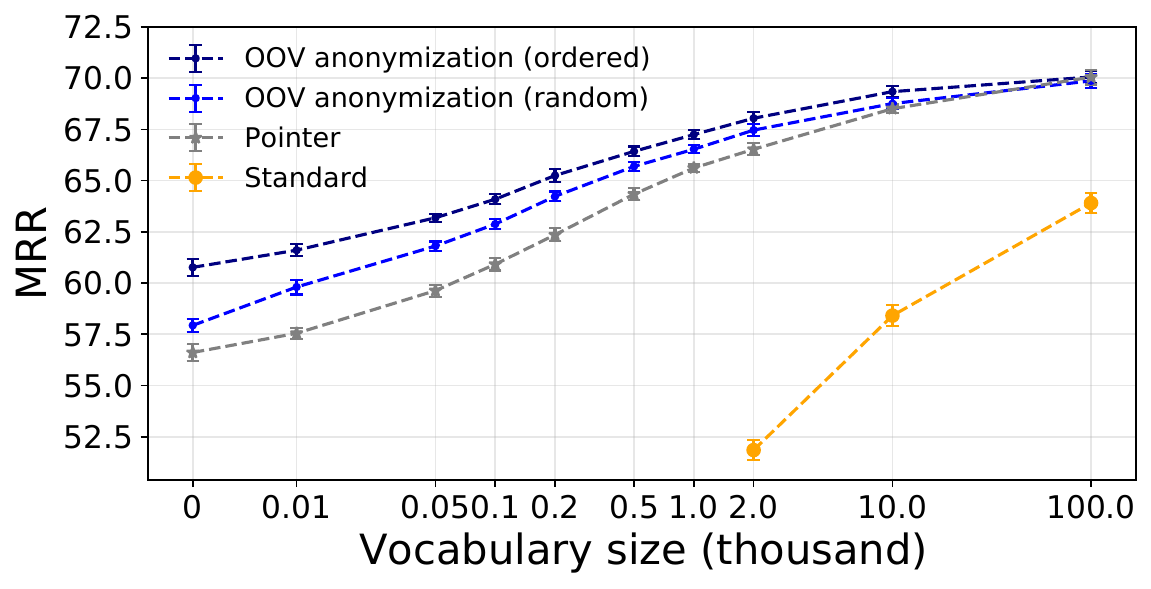} 
        \cprotect\caption{Results for Transformer in the code completion task (value prediction): mean reciprocal rank $\pm$ standard deviation over 3 runs. 
        The numerical data for the plots is given in Table~\ref{tab:results_all_cc} in Appendix.
        }
       \label{fig:anonymization}
\end{figure}

\textbf{Code completion task.\quad}
For the CC task, we use the setup of~\citet{code-prediction-transformer}, and below we briefly review it.
The CC task implies predicting the type and value of the next node based on the prefix of the depth-first AST traversal.
We predict the next type and value using two fully-connected heads on top of the Transformer decoder and optimize the sum of cross-entropy losses for types and values. While computing the loss, we skip the first occurrences of anonymized values and special positions, i.\,e.\,\verb|UNK| and \verb|PAD|. We tie the embeddings of input and output layers. In this task, we split the large AST traversals into chunks with a maximum length of $500$, as described in \cite{code-prediction-transformer}. The resulting dataset includes  186K~/~100K  training/testing chunks for Python and 270K~/~220K for JavaScript. 

We use mean reciprocal rank (MRR) to measure the quality of the model: 
$\mathrm{MRR} = 100\% \cdot N^{-1} \sum_{i=1}^N 1/\mathrm{rank}_i$, where $N$ is the total number of tokens in the dataset and $\mathrm{rank}_i$ is a position of the true token in the model ranking.
We assign zero scores (a) if the correct token is not in the top $10$ predicted tokens, (b) if the correct token is a \verb|UNK| and (c) for the first occurrences of anonymized identifiers.

For the next value prediction task, we add the pointer mechanism to the Transformer for comparison. We re-implement the pointer mechanism following the design choice of \cite{transformer_pointer}. Given an input sequence $[x_1, \ldots, x_\ell]$ of length $\ell$, Transformer outputs two distributions: the distribution over the fixed vocabulary $V$, $p_{\mathrm{model}}(a), a \in V$, and the probability of copying an input from position $j$, $p_{\mathrm{copy}}(j), j = 1, \ldots, \ell$. Then both distributions are combined to obtain the final distribution over the extended vocabulary:
$
    p(x_{\ell+1} = a) = p_{\mathrm{gen}} p_{\mathrm{model}}(a)[a \in V] + (1 - p_{\mathrm{gen}}) \sum_{j=1}^{\ell} p_{\mathrm{copy}}(j)[x_j = a] 
$. The switcher is computed given the current input and the output of the decoder as $p_{\mathrm{gen}}(x_\ell, h_\ell) = \sigma(w_h^T h_\ell + w_i^T x_\ell + b_{\mathrm{gen}})$. The cross entropy loss is computed over the extended vocabulary.

\subsection{Results}
We compare the proposed anonymization of OOV identifiers with the following baseline approaches: (1) Standard: with all OOV identifiers being replaced with an \verb|UNK| identifier; (2) training on fully anonymized data, i.\,e.\,
all identifiers are anonymized. This baseline corresponds to the zero vocabulary size in all plots. For the code completion task, we also include the baseline with the pointer mechanism.

Figure~\ref{fig:res_varmisuse} presents the results for the variable misuse task, for different frequent identifier vocabulary sizes. We observe that the proposed approach, with the anonymization of OOV identifiers (dark blue and blue lines), performs substantially better than the baseline models, particularly than the standard approach with OOV identifiers being replaced with an \verb|UNK| identifier (orange line). The leftmost point in both blue lines corresponds to the full anonymization baseline (zero vocabulary size). The ordered OOV anonymization (dark blue line) performs slightly better or similarly to the randomized OOV anonymization (blue line). We also experimented with the frequency-based OOV anonymization, i.\, e.\,sorting rare identifiers by frequencies in the code snippet and assigning \verb|VAR1| to the most frequent one, \verb|VAR2| to the next one etc. We found that such a strategy achieves the same quality as the ordered anonymization.

Increasing the vocabulary size for the standard model does not help much and even hurts the performance, i.\,e.\,the standard model with a vocabulary of 50K identifiers outperforms the one with the largest possible vocabulary.
The reason is that the embeddings of rare identifiers are updated only several times during the training and do not change a lot after being initialized randomly. On the contrast, anonymized identifiers occur quite frequently in the data, e.\,g.\,thousands of times, so their embeddings are updated regularly. As a result, it is more beneficial to anonymize rare identifiers than to include them in vocabulary.

The intuition behind why OOV anonymization performs well is that it saves information about variable repetition and thus does not change the algorithm that the code snippet implements. For example, in the buggy snippet \verb|with open(myfnm) as myfp:| \verb|data = myfnm.read()| (should be \verb|myfp.read()|), the model with OOV anonymization detects that OOV variables after \verb|as| and before \verb|read| are different and correctly predicts the bug, while the model with OOVs replaced with \verb|UNK| does not distinguish variables \verb|myfnm| and \verb|myfp| and cannot detect the bug.

Figure~\ref{fig:anonymization} presents the results for code completion (value prediction), for different frequent identifier vocabulary sizes.
In this task, the ordered OOV anonymization again slightly outperforms the randomized OOV anonymization, and they both substantially outperform the standard baseline and the baseline with full anonymization.
Moreover, the proposed OOV anonymization surpasses the strong pointer baseline for almost all vocabulary sizes. 
The advantage of the proposed OOV anonymization approach is that it helps the Transformer to distinguish OOV identifiers in the input, while the pointer mechanism enhances only the output layer. Also, in contrast to the pointer mechanism, OOV anonymization is much easier to implement. The pointer mechanism and the OOV anonymization could be straightforwardly combined, however, in our experiments, this combination  did not increase the scores compared to the maximum score of the OOV anonymization and the pointer. The results for type prediction are relatively the same as for the value prediction and can be found in Appendix~\ref{sec:appendix_types}. We visualize the t-SNE representations of the learned embeddings in Appendix~\ref{sec:appendix_embeddings}. 

\subsection{The influence of the anonymized vocabulary size}
\label{sec:additional_exps}
The randomized OOV anonymization strategy comprises the hyperparameter $|V_{an}|$, i.\,e.\,the size of the anonymized vocabulary. It should not be less than the maximum sequence length, to avoid using the same placeholder for different identifiers, and we select $|V_{an}|$ as the maximum length of code snippets. We tried using the larger values of $|V_{an}|$ and observed the insignificant difference in quality in the variable misuse task, and a slight drop in quality in the code completion task, as shown in Table~\ref{tab:anon_vocab}.

\begin{table}[!ht]
    \centering
    \begin{tabular}{c|c|c|c|c}
             & \multicolumn{2}{c|}{\textbf{PY}}   &  \multicolumn{2}{c}{\textbf{JS}} \\ \hline
          $|V_{full}|:$      & 1k      & 10k     & 1k        & 10k    \\ \hline
         $|V_{an}| = 0.5\mathrm{K}$    &  63.35 & 64.77  &  66.63  & 68.98 \\ 
         $|V_{an}| =1\mathrm{K}$   &  63.03 & 64.63  &  66.52  & 68.75 \\ 
         $|V_{an}| = 3\mathrm{K}$   &  62.79 & 64.34  &  66.22  & 68.60 \\ \hline
    \end{tabular}
    \cprotect\caption{Increasing the size $|V_{an}|$ of the anonymized vocabulary for two frequent identifier vocabulary sizes $V_{full}$, namely 1k and 10k, in the code completion task (value prediction). Metric: MRR (\%), all standard deviations are less than 0.3\%.}
    \label{tab:anon_vocab}
\end{table}

\section{Conclusion}
In this work, we propose the effective anonymization-based encoding of out-of-vocabulary identifiers, with two options, namely ordered and randomized OOV anonymization. Our preprocessing technique allows for easy implementation, could be easily plugged into various Transformer models and outperforms the widely used standard approach by a significant margin. The ordered anonymization performs slightly better than the randomized anonymization but requires a more careful implementation.

\section*{Acknowledgments}
We would like to thank Ivan Rubachev and the anonymous reviewers for the valuable feedback. The results on the variable misuse task 
were supported by the Russian Science Foundation grant \textnumero 19-71-30020.
The results on the code completion task 
were supported by Samsung Research, Samsung Electronics. The research was supported in part through the computational resources of HPC facilities at NRU HSE.

\bibliography{anthology,custom}
\bibliographystyle{acl_natbib}

\clearpage
\appendix

\section{Implementation details}
\label{sec:appendix_details}
\paragraph{Passing ASTs to Transformer.} To pass an AST to the Transformer, we follow the strategy of~\citet{chirkova2020empirical}.
They converted each input code snippet to the depth-first traversal of the abstract syntax tree (AST), obtaining a sequence of pairs (node type, node value).
The node types denote syntactic units of the programming language, e.\,g.\,\verb|If| or \verb|For|, and come from a small dictionary (up to 350 types), while the node values denote user-defined identifiers, language-specific identifiers, e.\,g.\,\verb|None| in Python, and constants. Some nodes do not store any values, we insert \verb|<EMPTY>| values in these nodes. We store two embedding layers, one for types and one for values, and sum the embedding of type and value in each AST node. The OOV anonymization is applied to the values. To train a model on the fully anonymized data, we anonymize all values except \verb|<EMPTY>|.

\textbf{Hyperparameters.} We list hyperparameters for the VM / CC tasks using slashes. Our Transformer model has 6 layers, $8$~/~$8$ heads, $d_{\mathrm{model}}$ equals to $512$~/~$512$. The number of parameters of our models (excluding embeddings) is 19M / 18M. We train all Transformers using Adam with a starting learning rate of 0.00001~/~0.001 and the batch size of 32 for 20 epochs (CC), 25 epochs (VM PY), or 40 epochs (VM JS). In the CC task, we use cosine learning rate schedule \cite{cosine} with a warmup step of 2000 and zero minimal learning rate, and the gradient clipping of 0.2. In the VM task, we use a constant learning rate. We use residual, embedding and attention dropout with $p=0.2~/~0.1$. We use relative attention~\cite{seq_rel_attn} with the maximum distance between elements of 8~/~32. 

\section{Validating our implementation}
\label{app:reproduce}
The numbers reported in our paper are not directly comparable to the works we borrow setups from, because we use our custom (and more correct) data split rather than the commonly used split (see details in Sec.~\ref{sec:experiments}). We ensure the validity of our results in two ways: by relying on the code of recently published works, and by comparing our numbers achieved for the commonly used data split to the numbers in the corresponding papers. Particularly, we use the model / loss / metrics and the code of \citep{code-prediction-transformer} for the CC task, and the model / loss / metrics of \citep{Hellendoorn} for the VM task (we rewrite line-by-line their code for metrics and loss). 
For the vanilla Transformer in the VM task, \citet{Hellendoorn} report 67.7\% joint accuracy and we achieved 64.4\% with the similar model size. The results are close to each other. 
For the vanilla Transformer in the CC task (Python), for value~/~type prediction, \citep{code-prediction-transformer} report 58.0 / 87.3 MRR (“TravTrans” model), and we achieve 60.0 / 89.1 MRR, again the results are close.

\section{Experiments with type prediction}
\label{sec:appendix_types}
\begin{figure}[!ht]
    \centering
    ~~~~~~Python150k dataset (custom train-test split) \\
         \includegraphics[width=\linewidth]{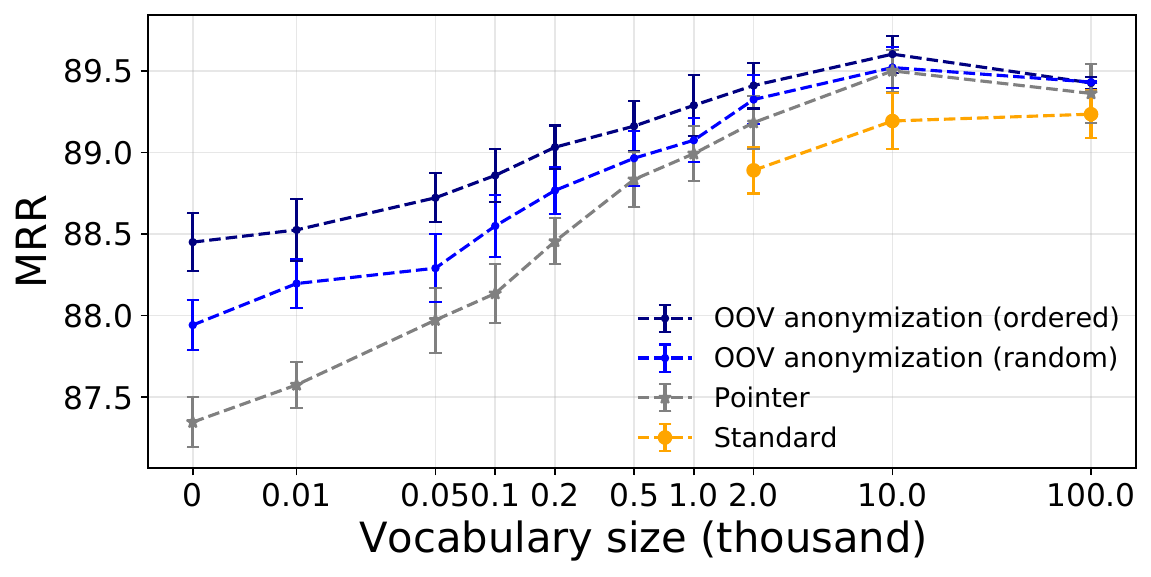}  \\
         ~~~JavaScript150k dataset (custom train-test split) \\
         \includegraphics[width=\linewidth]{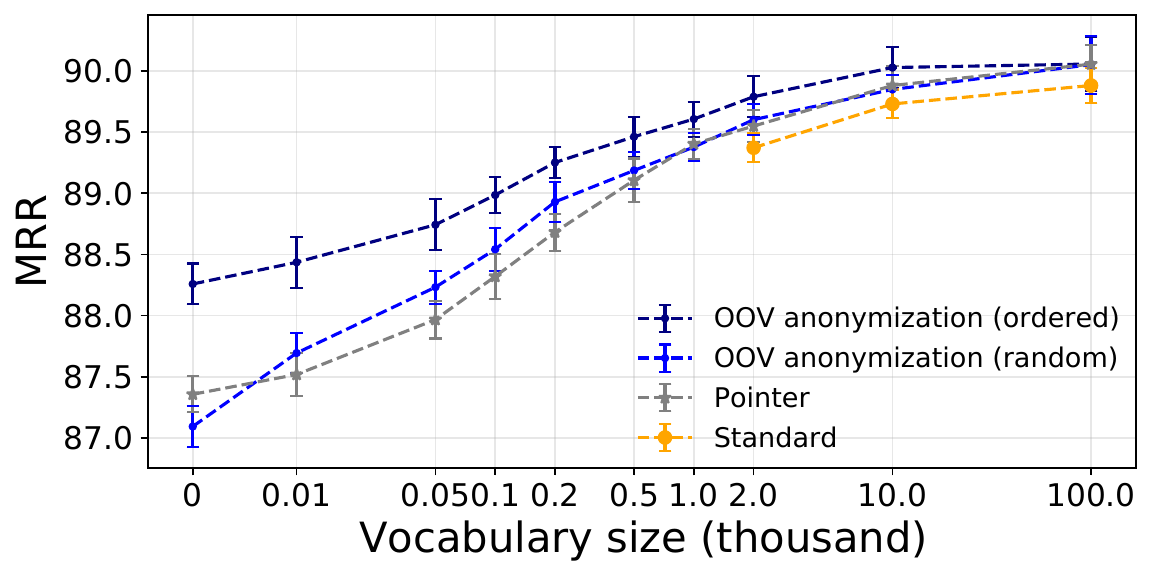} 
        \cprotect\caption{Type prediction  for code completion task for Transformer: mean reciprocal rank $\pm$ standard deviation over 3 runs.}
       \label{fig:types}
\end{figure}

In Fig.~\ref{fig:types}, we report the results for type prediction, code completion task.  Overall, the anonymization of rare identifiers again performs better, compared to the standard Transformer with rare identifiers replaced with an \verb|UNK|, and also improves over the pointer baseline for almost all vocabulary sizes.

\section{Visualization of embeddings}
\label{sec:appendix_embeddings}
\begin{figure*}[ht!]
    \centering
    ~~~~~~Python150k embeddings (custom train-test split) \\
        \centering
         \includegraphics[width=\linewidth]{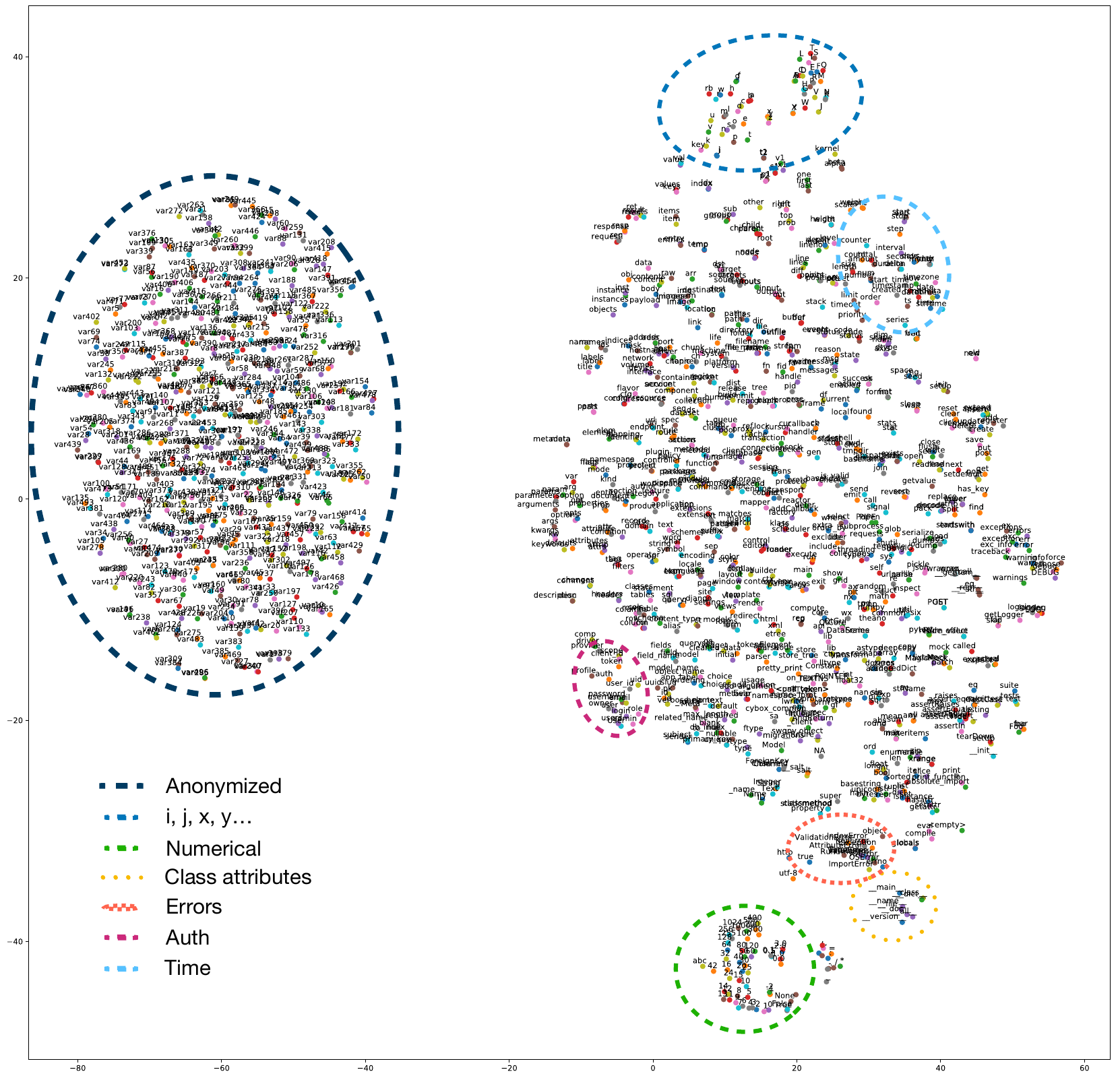}  \\
        \cprotect\caption{t-SNE visualization of Python150k  embeddings. Anonymized (random) embeddings (left cluster), in-vocabulary embeddings (right cluster).}
       \label{fig:embeddings}
\end{figure*}

In this section, we visualize the embeddings learned in the code completion task on Python dataset, vocabulary size 1k, OOV identifiers are anonymized randomly.
We use t-SNE \citep{tsne} with default hyperparameters and cosine metric to visualize the embeddings in a 2-dimensional space, see Figure ~\ref{fig:embeddings}. We observe that the embeddings of anonymized identifiers form a well-separated cluster in the embedding space. We also measured the inter-cluster~/~intra-cluster cosine similarities: the average cosine similarity between pairs of embeddings in one cluster~/~two different clusters. We observe the inter-cluster similarity for in-vocabulary~/~OOV identifiers of 0.09~/~0.05, and the intra-cluster similarity between in-vocabulary~/~OOV clusters of $-0.06$, which shows that
these two clusters occupy different subspaces of the embedding space.

\section{Numerical data for the plots in the paper}
Table~\ref{tab:results_all_vm} lists the numerical data for Figure~\ref{fig:res_varmisuse} and Table~\ref{tab:results_all_cc} lists the numerical data for Figure~\ref{fig:anonymization}.

For the code completion task, we also report accuracy scores of the best performing models for values prediction. We mark \verb|UNK| prediction as wrong. Our random / ordered / pointer / baseline models achieve $59.31$ / $59.71$  / $58.88$ / $50.41$ accuracy (\%) on the Python dataset, and  $64.08$ / $64.13$  /  $63.58$ / $58.48$ on the JavaScript dataset.

\clearpage

\begin{table*}[h!]
    \centering
    \begin{tabular}{c|c|c|c|c|c|c}
    & \multicolumn{6}{c}{Variable Misuse task, Joint localization and repair accuracy (\%)}  \\ \hline
    & \multicolumn{3}{c|}{PY}   &  \multicolumn{3}{c}{JS} \\ \hline
    $|V_{full}|$ &  Ordered &  Random &  Standard &  Ordered &  Random &  Standard \\ \hline
    1       &       81.93 &      81.26 &         0.00 &       78.80 &      77.69 &         0.00 \\
20      &       84.08 &      83.72 &        40.28 &       79.45 &      78.69 &         6.89 \\
100     &       85.29 &      84.96 &        54.71 &       81.63 &      80.40 &        33.66 \\
300     &       86.26 &      85.33 &        64.72 &       82.01 &      80.47 &        50.61 \\
1000    &       86.76 &      86.10 &        72.36 &       83.05 &      82.62 &        62.87 \\
10000   &       87.72 &      86.66 &        79.36 &       82.68 &      82.80 &        73.36 \\
25000   &       87.29 &      86.85 &        81.01 &       82.66 &      82.65 &        74.27 \\
50000   &       87.02 &      86.82 &        81.42 &       82.16 &      82.62 &        76.33 \\
Max &       N/A &        N/A &        80.16 &       N/A &      N/A &        73.68 \\
\end{tabular}
    \caption{Numerical data for Figure~\ref{fig:res_varmisuse}. Max denotes the vocabulary without any filtering: 622K for PY and 266K for JS. The standard deviations for all models are approx. 0.5\%.}
    \label{tab:results_all_vm}
\end{table*}

\begin{table*}[h!]
    \centering
    \begin{tabular}{c|c|c|c|c|c|c|c|c}
                    & \multicolumn{8}{c}{Code Completion task, Values Prediction, MRR (\%)}  \\ \hline
                    & \multicolumn{4}{c|}{PY}   &  \multicolumn{4}{c}{JS} \\ \hline
    $|V_{full}|$          &      Ordered   &   Random &  Pointer & Standard                 & Ordered    & Random  & Pointer  & Standard \\ \hline
        1                 &          58.61 &   57.00 &    54.77 &      N\textbackslash A    &     60.77  & 57.93   &    56.61 &      N\textbackslash A \\
        10                &          59.52 &   58.40 &    56.01 &      N\textbackslash A    &     61.60  & 59.81   &    57.55 &      N\textbackslash A \\
        50                &          60.74 &   59.45 &    57.68 &      N\textbackslash A    &     63.19  & 61.81   &    59.63 &      N\textbackslash A \\
        100               &          61.41 &   60.35 &    58.53 &      N\textbackslash A    &     64.09  & 62.88   &    60.92 &      N\textbackslash A \\
        200               &          62.26 &   61.31 &    59.72 &      N\textbackslash A    &     65.25  & 64.23   &    62.36 &      N\textbackslash A \\
        500               &          63.29 &   62.55 &    61.45 &      N\textbackslash A    &     66.43  & 65.69   &    64.34 &      N\textbackslash A \\
        1000              &          63.90 &   63.12 &    62.38 &      N\textbackslash A    &     67.24  & 66.53   &    65.61 &      N\textbackslash A \\
        2000              &          64.51 &   64.05 &    63.28 &     46.0                  &     68.04  & 67.46   &    66.53 &    51.85 \\
        10000             &          65.25 &   64.77 &    64.94 &     51.8                  &     69.34  & 68.76   &    68.51 &    58.41 \\
        100000            &          64.27 &   63.91 &    65.09 &    55.07                  &     70.05  & 69.87   &    70.05 &    63.9 \\
    \end{tabular}
    \caption{Nu   merical data for Figure~\ref{fig:anonymization}. The standard deviations for all models are approx. 0.3\%.}
    \label{tab:results_all_cc}
\end{table*}

\end{document}